# Microstructural and dielectric properties of $Ba_{0.6}Sr_{0.4}Ti_{1-x}Zr_xO_3$ based combinatorial thin film capacitors library

Guozhen Liu[a)], Jerome Wolfman [b)], Cécile Autret-Lambert, Joe Sakai, Sylvain Roger, Monique Gervais, François Gervais

*Université François Rabelais, Laboratoire d'Electrodynamique des Materiaux Avancés (LEMA), UMR 6157 CNRS/CEA, Parc de Grandmont, 37200 Tours, France*

**Abstract**

Epitaxial growth of $Ba_{0.6}Sr_{0.4}Ti_{1-x}Zr_xO_3$ ($0 \leq x \leq 0.3$) composition spread thin film library on $SrRuO_3/SrTiO_3$ layer by combinatorial pulsed laser deposition (PLD) is reported. X-ray diffraction and energy dispersive x-ray spectroscopy studies showed an accurate control of the film phase and composition by combinatorial PLD. A complex evolution of the microstructure and morphology with composition of the library is described, resulting from the interplay between epitaxial stress, increased chemical pressure and reduced elastic energy upon Zr doping. Statistical and temperature-related capacitive measurements across the library showed unexpected variations of the dielectric properties. Doping windows with enhanced permittivity and tunability are identified, and correlated to microstructural properties.

**a) present address :** Department of physics, Temple university, Philadelphia, PA 19122, USA

b) Corresponding author : wolfman@univ-tours.fr

# 1. Introduction

$Ba_{1-x}Sr_xTiO_3$ (BST) solid solutions have been widely studied due to their potential applications in tunable microwave devices and dynamic random access memories.[1] BST ferroelectric-paraelectric Curie temperature transition, and thus room temperature (RT) permittivity, can be tailored over a broad range by changing the Sr concentration.[2,3] Moreover, the permittivity of BST exhibit a large nonlinear dependence on the applied electric field around the Curie temperature.[4] These properties make BST an excellent starting point to develop new materials that meet various electronic components requirements. Recently, $BaTi_{1-x}Zr_xO_3$ (BTZ) solid solutions have attracted much attention because of their high dielectric tunability and low loss.[5,6] In addition, it is also reported that $Zr^{4+}$ substituting for $Ti^{4+}$ in $BaTiO_3$ can result in low and stable leakage current since $Zr^{4+}$ is chemically more stable than $Ti^{4+}$.[7,8] However, the permittivity of BTZ is significantly reduced with respect to that of BST. Doping $BaTiO_3$ with both Sr and Zr is the natural next step. So far, few reports on the combined effects of Sr and Zr doping in $BaTiO_3$ have been published and most of them dealed with ceramics or polycrystalline films on Si wafers.[9,10,11,12]

Change in the composition leads to cell parameters variation, which in the case of epitaxial thin films results in various distortions. It is well known that titanate dielectric properties are very sensitive to distortions [4,13]. Systematic studies on the combined effects of Sr and Zr doping in $BaTiO_3$ epitaxial thin films are therefore needed.

In this paper, we present a systematic study of Zr doping effects on the microstructural and dielectric properties of an epitaxial $Ba_{0.6}Sr_{0.4}TiO_3$ thin film library deposited onto $SrRuO_3$ (SRO) bottom electrode. The continuous composition-spread $Ba_{0.6}Sr_{0.4}Ti_{1-x}Zr_xO_3$ (BSTZ-$x$ with $0 \leq x \leq 0.3$) thin film was prepared by combinatorial pulsed laser deposition. After Au top electrode shadow-mask deposition, this library of 160 capacitors offered the unique ability, although seldom reported in the literature,[14,15] to statistically explore the effect of the interplay between doping and substrate induced strain on the dielectric properties. Both the microstructure and dielectric properties of the BSTZ-*x* thin film showed non-monotonic dependences on Zr concentration. Possible origins of the composition-dependent behaviors are discussed.

## 2. Experimental procedure

Epitaxial BSTZ-*x* thin film library and SRO layer were grown by pulsed laser deposition on (001) SrTiO$_3$ (STO) 10 mm x 10 mm substrates. A KrF excimer laser ($\lambda$ = 248 nm) with a fluence of 2 J/cm$^2$ was used to ablate ceramic targets synthesized by an organic gel-assisted citrate process.[16] After heating the substrate up to 600$^o$C under vacuum, a 60-nm-thick SRO layer (lattice parameters a=5.57Å, b=5.53 Å, and c=7.85 Å) was deposited at a laser repetition rate of 5 Hz in a dynamic oxygen pressure of 5 Pa. The oxygen pressure was then increased to 10 Pa, maintaining the substrate temperature. The 400-nm thick BSTZ-*x* films were deposited using two ceramic targets, namely BSTZ-0 and BSTZ-0.3, having a deposition rate of about 25 pulses per unit cell (u.c.). A computer controlled translating mask (x direction) with a rectangular aperture was inserted close to STO substrate (about 0.5 mm) to localize the deposition. Synchronizing the mask movements and the laser firing on the first target, we exposed one side of the substrate for 25 laser pulses (i.e. a complete u.c.) while the other side is exposed for only one laser pulse, with intermediate pulse number exposure along intermediate substrate locations. This way a single layer with a varying degree of completeness is deposited. Selecting the second target and using the complementary sequence of mask positions and laser firing, a complete monolayer is obtained whose composition varies from BSTZ-0 to BSTZ-0.3 in the x direction, while the composition is kept constant in the y direction. The 400-nm thick BSTZ-*x* films were obtained by repeating the previous sequence 1000 times. A schematic of the synthesized samples is shown Fig. 1, and more details about the combinatorial growth can be found in ref. 14.

The local compositions were mapped by energy dispersive x-ray spectroscopy (EDX) using an electron beam probe perpendicular to the surface. Local phase analysis was conducted using a prototype high throughput X-ray micro-diffractometer (Bruker-AXS D8 discover HTS) in $\theta$-$2\theta$ scans mode. In this micro-diffractometer, the beam is focused at the sample position (footprint 200 µm x 50 µm at $\theta$ = 45°) and data are collected with a linear detector. This beam geometry, optimal for epitaxial film, gives however rise to peak enlargement for mosaic thin films. The structural characterization of epitaxial films was undertaken with reciprocal space mapping (RSM) around the (103) STO reflection using a high resolution X-ray diffractometer (Bruker-AXS D8 discover with Cu $K_{\alpha 1}$ parallel beam). Measurements were done for steep incidence of the X-ray beam with a 50µm-width slit on the primary optic side to limit the beam footprint on the sample to a 92µm-width strip along the y direction. RSM were acquired every mm in the x direction. Surface morphology observations and lamella preparation for transmission electron microscopy (TEM) were done with a dual beam FIB-SEM FEI Strata 400. High resolution TEM imaging was realized using a JEOL FEG 2100. For electrical measurements, rectangular Au top electrodes (150 µm along x axis and 450 µm along y axis) were sputter deposited through a shadow mask. Complex impedance versus temperature was measured by a HP4294A impedance analyzer with a 30 mV oscillation voltage.

## 3. Results and discussions

Figure 2 shows a contour plot of the Zr / (Zr + Ti) composition ratio across the BSTZ-$x$ thin film library. Zr concentration varies continuously and almost linearly along the gradient direction x. Vertical contour lines, with a spacing of 3-4 at. %, are observed along the nominally homogeneous direction. The undulations of the contour lines are believed to result from the limited EDX precision (about 2-3 at.% due to signal convolution from the STO substrate) rather than to local composition fluctuations. The measured Zr concentrations, close to 0% at one side and 30% at the other side, are consistent with the nominal values, evidencing a good control of local composition in the combinatorial deposition set-up.

Figure 3a shows $\theta$-$2\theta$ X-ray diffraction (XRD) patterns recorded every mm along the library x axis, corresponding to a nominal increment in Zr content of $\Delta x = 0.033$. A blow-up of these patterns around $2\theta = 45.5°$ is presented in Fig. 3b. All the XRD patterns evidenced strong (00l) perovskite peaks with no secondary phases (Fig. 3a). On the left of STO (002) peak (Fig. 3b at $2\theta = 46.49°$), a peak compatible with an evolving BSTZ-x (002) moves from $2\theta = 44.6°$ ($x = 0.3$) to $2\theta = 45.1°$ ($x = 0$). SRO single layer deposited on STO evidenced a (002) SRO peak at $2\theta = 45.2°$ (not shown), which makes it difficult to distinguish from the BSTZ-x (002) peak. Some extra contribution to the intensity around $2\theta = 45.7°$ is visible in Fig. 3 for $x \in [0.07; 0.21]$ with a maximum at $x = 0.18$. It is not possible to attribute this extra intensity neither to BSTZ-x nor to SRO peaks from $\theta$-$2\theta$ geometry. RSMs around (103) STO reflection were acquired every mm along the library x axis to further investigate. Fig. 4 shows some of these BSTZ-x/SRO/STO stack RSMs for $x = 0, 0.1, 0.2$ and $0.3$. The substrate peak is obviously the most intense at h = 1 and l = 3. The peak located at h = 1 and elongated mainly along l direction around l = 2.92 corresponds to a coherently strained layer with a varying out of plane parameter, and is assigned to $(103)_{SRO}$. The peak with the smallest h and l values comes from a relaxed layer with the largest in-plane and out-of-plane parameters. Considering the bulk parameters of BSTZ-x[11], this last peak is assigned unambiguously to $(103)_{BSTZ-x}$. The proximity of SRO, BSTZ-x, and STO (103) lattice points indicates that SRO and BSTZ-x are epitaxial onto STO, with a cube on cube type of growth. The dark dotted lines in Fig. 4 are stretching from $(103)_{STO}$ lattice point to the reciprocal space origin. All (103) lattice points originating from cubic cells should lie on this line. It can be seen from Fig. 4 that $(103)_{BSTZ-x}$ peak is slightly bellow the line for $x = 0.3$, and on the line for $x = 0.1$ and $0.2$, therefore the films are partially strained for $x = 0.3$ and fully relaxed for $x = 0.1$ and $0.2$, assuming a cubic paraelectric BSTZ-$x$ at RT. In the case of $x = 0$, the (103) peak is also slightly bellow the line (Fig. 4), but we shall see later on that this tetragonal distortion is to be related to a stress induced ferroelectric state at RT. The in-plane and out-of-plane lattice parameters a and c have been extracted from (103)$_{BSTZ-x}$ peak maximum positions, and are plotted versus $x$ in Fig. 5a. In-plane parameter *a* exhibits a plateau at about 0.4015 nm up to $x = 0.1$, then increases toward 0.4026 nm where another plateau occurs between $x = 0.17$ and $x = 0.27$ before increasing again to 0.4032 nm for $x = 0.3$. Out of plane parameter c remains almost unchanged at about 0.4028 nm up to $x = 0.17$ and then steadily increases up to 0.4051 nm for $x = 0.3$, consistent with XRD results in $\theta$–$2\theta$ geometry (Fig. 3). It can be seen that BSTZ-$x$ is fully relaxed, i.e. a = c within the experimental error, for $x \in [0.1; 0.2]$, while c parameter is elongated for $x < 0.1$ and $x > 0.2$. The $Ba_{0.6}Sr_{0.4}TiO_3$ lattice parameters reported here (a = 0.4016 nm and c = 0.4028 nm) differ from those reported in the literature. Ohtani et al. reported parameters a = 0.396 nm and c = 0.405 nm for $Ba_{0.6}Sr_{0.4}TiO_3$ of undisclosed thickness grown by PLD on STO[17], while Abe et al. reported a = 0.391 nm and c = 0.426 nm for 100 nm thick $Ba_{0.6}Sr_{0.4}TiO_3$ grown onto SRO/STO by RF sputtering[18]. However it has been shown that, in the $Ba_{1-x}Sr_xTiO_3/SrTiO_3$ system, x = 0.6 is the threshold at which compressive stress is abruptly relaxed through the introduction of misfit dislocations,[17] leading to rapid change of in-plane and out of plane lattice parameter. Slight variations of the Ba/Sr ratio, and/or of the relaxation mechanism among the films could explain such a parameter distribution in the literature. It should be underlined that in the BSTZ-$x$ system almost all lattice parameters reported in the literature for relaxed films are larger than their bulk counterpart, as it is the case here. Oxygen deficiency is the most commonly cited hypothesis to explain this fact, being at the same time the most probable and the most difficult to check in films. Another important factor is $Ba_{0.6}Sr_{0.4}TiO_3$ Curie temperature which is closed to 0 °C in ceramics[11] and may be shifted above RT in stressed thin films,[4] leading to a tetragonal distortion. The structure distortion, i.e. the a/c ratio, is plotted versus $x$ in Fig. 5b (black squares). It increases from 0.997 to 1.001 as $x$ increases from 0 to 0.17 and then decreases to about 0.996 when $x$ reaches 0.3. In ceramic BSTZ-x solid solution, no distortion is observed up to x = 0.3.[11] Distortions observed here originates from epitaxial stress. Looking back at Fig. 4, the shape of (103)$_{BSTZ-x}$ peak evolves from rounded (x = 0) to triangular (x > 0). Two factors are known to promote such triangular peak shape, namely film relaxation and the correlated mosaicity[19]. BSTZ-x cell parameters are larger than STO[11] (Fig. 5a) so the corresponding films experience compressive stress from the substrate. Above a critical thickness relaxation occurs, leading to peak shifting along the RSM relaxation line, which is accompanied by peak shape deformation if multiple strain states are present. Cell parameters of a fully strained $Ba_{0.6}Sr_{0.4}Ti_{0.7}Zr_{0.3}O_3$ film, extracted from ref. 20, have been used to tentatively draw the relaxation line for $x = 0.3$ (light dotted line in Fig. 4). It has been shown that $Ba_{0.6}Sr_{0.4}Ti_{0.7}Zr_{0.3}O_3$ critical thickness on STO is about 10 nm.[20] As relaxation occurs above 10nm, misfit dislocations are formed and lead to mosaicity increase.[19] The peak deformation related to increased mosaicity occurs along a direction perpendicular to the scattering vector, i.e. the ω scan direction, represented in Fig. 4 by white dotted lines for $x = 0.2$. The conjunction of a peak progressive enlargement along ω scan direction while the peak is shifting along the relaxation line explains the observed triangular shape.[19] The rounded peak shape observed for $x = 0$ (Fig. 4) compared to triangular shapes otherwise indicates an abrupt relaxation in BSTZ-*0* at early growth stage, while various strain states coexist within the film thickness upon Zr doping. The mosaicity of BSTZ-x, SRO and STO layers and substrate was estimated from FWHM of section along ω scan direction (i.e. rocking curve RC) at the (103) peaks maxima within the RSMs. BSTZ-x mosaicity is plotted versus $x$ in Fig. 5b (open circles). RC FWHM measured for STO and SRO, below 0.03° and 0.07° respectively, were much smaller than that for BSTZ-x, so their impact on BSTZ-x mosaicity is negligible. Interestingly BSTZ-x RC FWHM and distortion exhibit very similar trends (Fig. 5b), underlining the correlation between relaxation and mosaicity. A mosaicity maximum of about 0.7° is observed for $x = 0.13$ corresponding to a fully relaxed layer. The lower crystalline disorder (i.e. mosaicity) observed at $x = 0.3$ compared to $x = 0$ is surprising as the mismatch with STO increased upon Zr doping[11]. This implies a lower misfit dislocation density and therefore a reduced elastic energy for $x = 0.3$. This deduction is consistent with the abrupt relaxation observed for $x = 0$ and the progressive one for $x = 0.3$ (Fig. 4). The continuous decrease of mosaicity from $x = 0.13$ to $x = 0.3$, as the mismatch increases, also pleads for a continuous reduction of stored elastic energy in BSTZ-x upon Zr doping. That is the competition between counteracting elastic energy reduction and increased mismatch upon Zr doping which is responsible for the pointed shape of the mosaicity versus $x$ curve (Fig. 5b). Turning back to SRO (103) peak shape in Fig. 4, we see that while it gets elongated toward higher l values as x increases, it also inflates and tilts towards lower h values. This indicates a small in-plane lattice relaxation which probably results from the growing stress exerted by BSTZ-*x* on SRO as *x* increases.

Figure 6 show four SEM images of BSTZ-*x* thin films for $x = 0, 0.03, 0.13$ and $0.3$, respectively. From the top-left image, it can be seen that BSTZ-*0* film is mainly composed of tapered protruding grains, about 100 nm long, with triangular shapes. As we have seen from X-ray diffraction that all BSTZ-*x* films are (001) oriented, we can say that {001} surface does not have the lowest surface energy for BSTZ-*0*. This situation however evolves upon Zr doping. For $x = 0.03$, flat surface appears in between tapered grains (Fig. 6), composed of smaller patchy square grains. As *x* increases, the volume fraction of tapered protruding grains decreased, and the volume fraction of small flat grains increased correspondingly. For the BSTZ-*0.3* thin film, the surface is largely dominated by small grains. Such a decrease in average grain size in BST upon Zr doping is also observed by other groups.[11,21] It appears that

{001} surface energy compares more and more favourably to other surfaces orientation energy upon Zr doping, as is commonly seen for perovskites.[22] To further characterize the film morphology and ascertain homogenous and unique composition within both types of grains, we extracted and thinned a lamella from BSTZ-*0.13* film location for TEM observation. Images of the lamella taken at various magnifications are presented in Fig. 7. Very dense layers are visible in Fig. 7a, with contrasts along the growth direction in both SRO and BSTZ-*0.13*, underlining a columnar growth. The diameter of the column is in the 20-40 nm range. Beside columnar grains, a few tapered grains protrude from the BSTZ top surface, and extend within the layer with cone shapes. These cones are unambiguously the immerged part of the triangular grains observed by SEM from above (cf. Fig. 6). At higher magnification (Fig. 7b) we see that the cones apexes are located at about 100 nm from the BSTZ-*0.13* / SRO interface. The sides of the cones seem to correspond to misfit dislocations. On STEM image (not shown) it appears that dislocations do not start at the cone apex but rather stretch out towards the BSTZ-*0.13* / SRO interface. One plausible hypothesis is that cone-like grains growth is initiated at misfit dislocations intersections. The continuously decreasing density of triangular protruding grain as *x* increases (see Fig. 6) would then be correlated to the reduction of misfit dislocations upon Zr doping. EDX mappings with electron probe size of a few nm showed no composition difference between columnar and cone shaped grains (not shown), with homogeneous composition within the grains. Furthermore, local Fourier transform of the images, equivalent to local electron diffraction, present the same pattern for areas located on each side of the misfit dislocation (Fig. 7c). This implies that crystalline orientation within cones and columns is the same, i.e. [00l].

Dielectric permittivity ($\varepsilon_r$), loss tangent (D) and tunabiliy ($n_r$) measured at 100 kHz across the BSTZ-x thin film library are presented in Fig. 8. Here the tunability is defined as $n_r = [\varepsilon_r(max)-\varepsilon_r(5V)]/\varepsilon_r(max)$ and the error bars relate to the statistical distribution observed for a given composition. It is clear from Fig. 8 that $\varepsilon_r$, D and $n_r$ show correlated dependences on Zr content, with two local maximum and one local minimum. A small shift is however observed between the first permittivity and tunability peak (0.07 and 0.1 respectively). These non-monotonic tunability and permittivity behaviors strongly contrast with results from BSTZ-x ceramics[11] and polycristalline thin films[12] where these two quantities continuously decrease upon Zr doping. Losses are also non-monotonic for bulk compounds,[11] but with a global minimum at $x = 0.15$ for $x \in [0 ; 0.3]$. Several effects can contribute to these unexpected variations of the dielectric properties in the case of epitaxial thin films. First the epitaxial stress leading to a strain along the probing electric field direction, i.e. compressive stress with an elongation along c in our MIM case, could increase the ionic displacement and thereby promote strain-induced polarization in the field direction, thus enhancing the dielectric constant and tunability.[13] Small tensile stress imposed by a thin $Ba_{0.7}Sr_{0.3}TiO_3$ buffer layer onto MgO substrate has indeed been reported to increase *in-plane* $Ba_{0.6}Sr_{0.4}TiO_3$ permittivity and tunability (inter-digital electrode IDE geometry).[23] Second, in the case of epitaxial $Ba_{0.6}Sr_{0.4}TiO_3$ films grown on LSAT with compressive stress, the significant reduction of the in-plane dielectric constant (IDE geometry) was attributed to local strain associated *with a large density of threading dislocations*.[24] It is interesting to note that the local minima observed around $x = 0.15$ for permittivity, tunability and losses (Fig. 8) correspond to a relaxed film (a = c in Fig. 5a) having the maximum mosaicity (Fig. 5b), i.e. the highest threading dislocation density. The last factor to take into account is a possible shift of the film permittivity maximum temperature $T_{max}$, due to the epitaxial stress from the substrate.[4] To explore such a possibility, capacitance versus temperature have been measured for several representative capacitors with various BSTZ-x compositions, and are represented in Fig. 9. $T_{max}$ versus $x$ is plotted in the inset of Fig. 9. The first striking observation is that $T_{max}$ first increases upon Zr doping, and then decreases at higher doping level. In bulk BSTZ-*x*, $T_{max}$ is reported to continuously decrease upon Zr doping, from 280K down to 270K, 235K and 175K for $x$ = 0, 0.1, 0.2 and 0.3 respectively[2,11]. Here we found that, for $x \in [0 ; 0.09]$, $T_{max}$ values are higher than for their bulk counterpart, and are even above RT for $x$ = 0 and $x$ = 0.03 (300K and 325K respectively). There is indeed a stress-induced $T_{max}$ shift toward higher temperature due to epitaxial growth and increased chemical pressure at small Zr doping. For higher Zr doping ($x \geq 0.21$), $T_{max}$ values are however smaller than bulk values.[11] Noticeably, maximum capacitances measured at $T_{max}$ continuously increases with $x$ (Fig. 9), contrary to what is observed for bulk[11]. For BSTZ-*0.21*, a rapid capacitance fall-off close to $T_{max}$ (44% drop over 20K, cf. Fig. 9) is followed by a much broader transition up to 370K. Hence the capacitance of BSTZ-*0.21* is still more than twice the capacitance of BSTZ-*0* at 320K (1.4 nF and 0.65 nF respectively). Temperature dependence of BSTZ-*0.21* tunability is illustrated in Fig. 10 where tunability versus electric field is plotted for various temperatures. It can be seen that the curves are not symmetrical (Fig. 10), the maximum permittivity being obtained here for an electric field of about -15 kV/cm. A systematic variation of this field offset has been observed across the library (not shown), which relates to the energy difference between electrodes work function and BSTZ-*x* electronic affinity, and will be discussed together with leakage current properties in a future publication. The maximum tunability, over 70% for 120 kV/cm, is obtained at 150K, i.e. at $T_{max}$ (Fig. 10). As the temperature increases from $T_{max}$ to RT, the permittivity slowly decreases, with a value of still about 50% at 290K. After these capacitance measurements versus temperature and electric field, several statements can be made concerning the correlation between Zr doping, structure and dielectric properties. First the RT permittivity and tunability enhancement observed at low Zr doping (first peak in Fig. 8) result, at least partly, from stress-induced $T_{max}$ increase and is associated to a tetragonal distortion. The following permittivity and tunability drop with a local minimum around $x$ = 0.15 correspond to a lowered $T_{max}$ and a full film relaxation, i.e. no distortion. The second permittivity and tunability peak (around $x = 0.23$ in Fig. 8) is not anymore associated with a transition at $T_{max}$ but rather with a broad transition in temperature (as exemplified for $x = 0.21$ in Fig. 9). It is associated with the return of the tetragonal distortion linked to the elongation of the c lattice parameter observed for $x > 0.17$ (Fig. 5b). Finally the decrease in tunability and permittivity observed at larger Zr doping (x > 0.23 Fig. 8) probably has chemical origin, i.e. the epitaxial film properties finally tend toward the bulk properties.

## 4. Conclusions

In conclusion, the effects of Zr doping on the microstructures and dielectric properties of composition spread BSTZ-$x$ ($0 \leq x \leq 0.3$) thin film based capacitors library were investigated. The interplay between epitaxial stress, increased chemical pressure and reduced elastic energy upon Zr doping has been shown to lead to a complex evolution of the film microstructure and morphology, correlated with non monotonic dielectric properties. Two doping windows with enhanced permittivity and tunability have been found. The first one around $x = 0.09$ is associated to a chemical and epitaxial stress-induced increase of the temperature $T_{max}$ close to or above RT leading to a tetragonal distortion. The second window, around $x = 0.23$, results from a broad ferroelectric-paraelectric transition extending well above RT from a much reduced $T_{max}$ and is also associated with a tetragonal distortion. Intermediate and higher Zr doping, around $x = 0.15$ and above $x = 0.23$ respectively, lead to a reduced permittivity and tunability. In the former case this reduction is associated with fully relaxed films having the highest mosaicity, while in the latter case bulk chemical properties finally overcome epitaxial influence. As a concluding remark, it has been shown that $BaSrTiO_3$ structural and dielectric properties are extremely sensitive to Zr doping level in a non-predictive way. A systematic and continuous phase diagram exploration is thus desirable to better optimize dielectric films for practical applications. Combinatorial deposition precisely provides a fast and very effective method to do so.

**Acknowledgments**

This work was supported by the French National Research Agency (ANR) through the NANOCOMBI project (Grant No. ANR-05-NANO-005-01) and by the Pole of competitiveness S2E2 through the SESAME-CAPI project.

Figure captions

Fig. 1 Schematic of the composition-spread BSTZ-$x$ thin films based capacitor library.

Fig. 2 (colored online) Zr concentration contour map across the BSTZ-$x$ library by EDX.

Fig. 3 (colored online) a) XRD $\theta$-$2\theta$ diffraction patterns of BSTZ-$x$ thin films with various Zr concentrations. b) zoom of Fig. 3a

Fig. 4 (colored online) XRD RSMs around (103) reflection in reciprocal space units of BSTZ-$x$ films on SRO/STO for $x$ = 0, 0.1, 0.2 and 0.3.

Fig. 5 (colored online) a) Lattice parameters ($a$, $c$) and b) lattice distortion ratio ($c/a$) and mosaicity of BSTZ-$x$ vs $x$.

Fig. 6 SEM images of BSTZ-$x$ thin films top surface for $x$ = 0, 0.03, 0.13 and 0.3

Fig. 7 TEM images of BSTZ-*0.13* thin film lamella at increasing magnification (a, b, c). Superimposed to the high resolution TEM image c) are the local image Fourier transforms showing the same orientation of the film on both side of the dislocation.

Fig. 8 (colored online) Zr concentration dependence of the permittivity $\varepsilon_r$, tunability $n_r$ and loss tangent $D$.

Fig. 9 (colored online) Capacitance vs temperature of BSTZ-x based capacitors for various Zr concentration. Inset : $T_{max}$ vs Zr concentration.

Fig. 10 (colored online) Electric field dependence on BSTZ-0.21 based capacitor tunability for various temperatures.

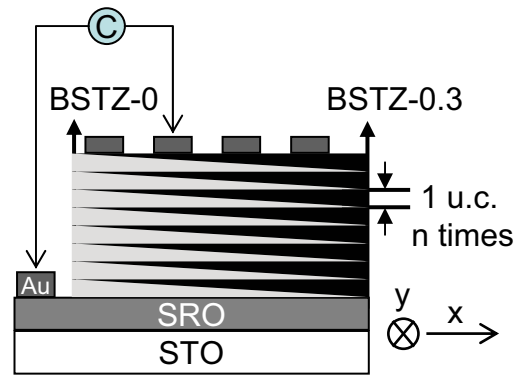

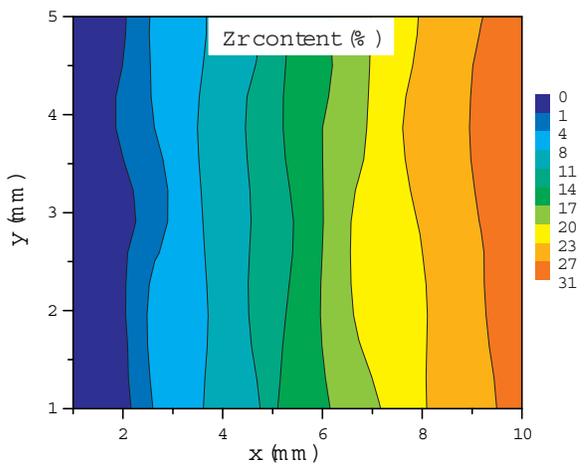

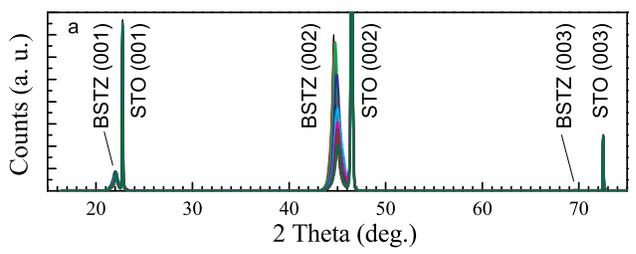

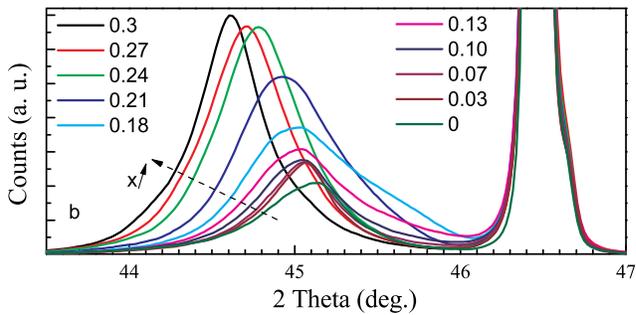

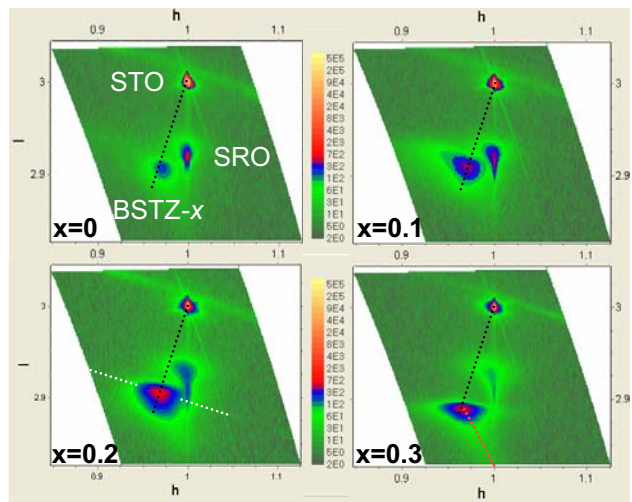

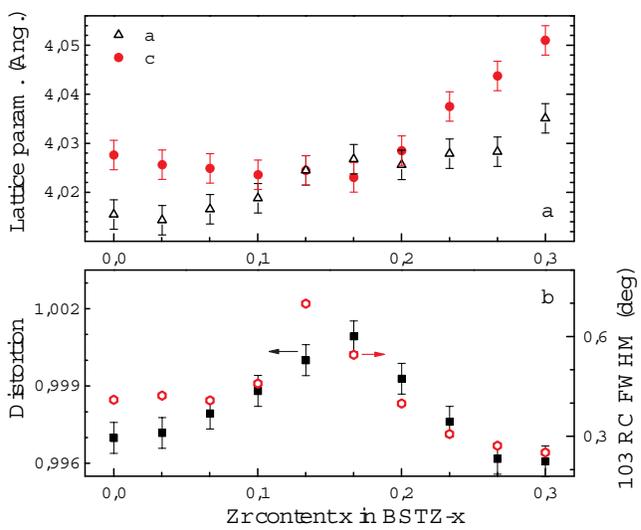

| x = 0 | x = 0.03 |
|---|---|
| 500nm | 500nm |
| x = 0.13 | x = 0.3 |
| 500nm | 500nm |

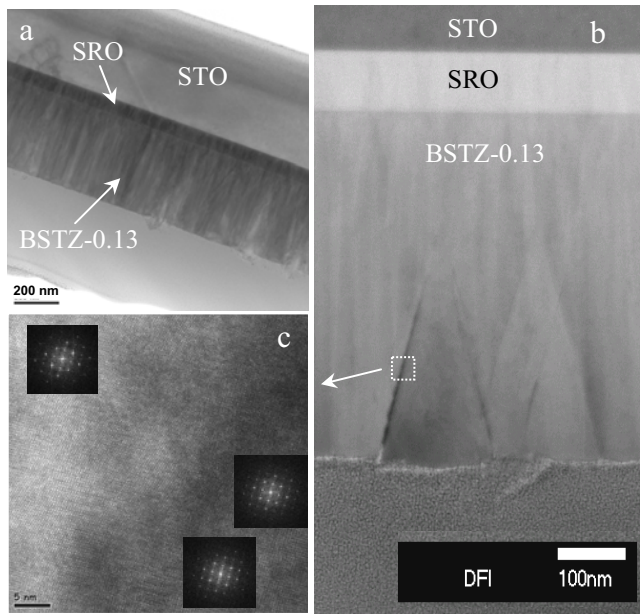

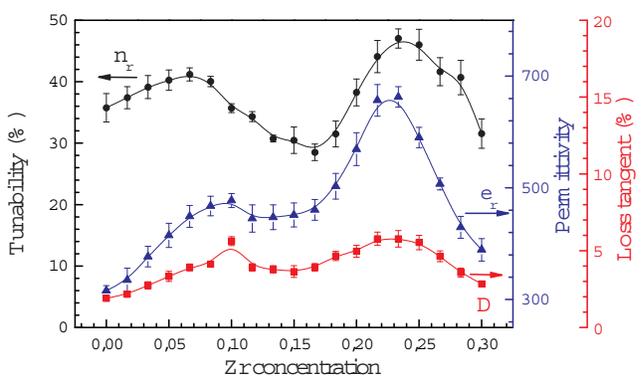

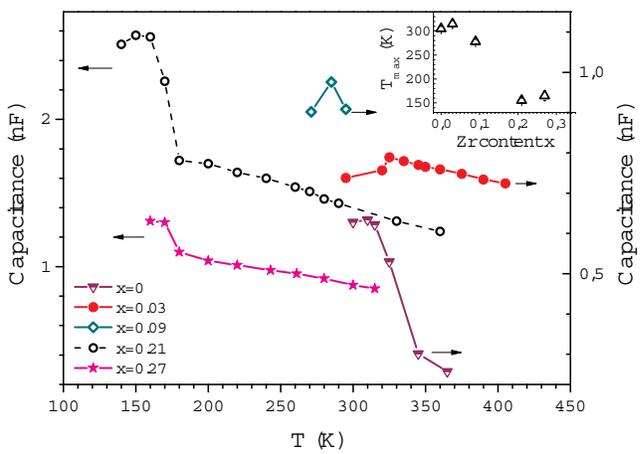

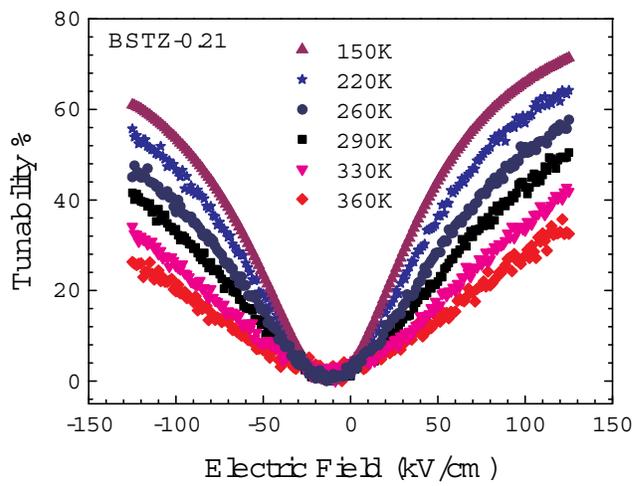